\documentclass{article}

\usepackage{arxiv}

\usepackage[utf8]{inputenc} 
\usepackage[T1]{fontenc}    
\usepackage[hidelinks,colorlinks=true,linkcolor=blue,citecolor=blue]{hyperref}       
\usepackage{url}            
\usepackage{booktabs}       
\usepackage{amsfonts}       
\usepackage{nicefrac}       
\usepackage{microtype}      
\usepackage{lipsum}		
\usepackage{graphicx}
\usepackage{natbib}
\usepackage{doi}

\title{CML-COVID: a large-scale COVID-19 Twitter dataset with latent topics, sentiment and location information}

\author{ \href{https://orcid.org/0000-0001-6400-1190}{\includegraphics[scale=0.06]{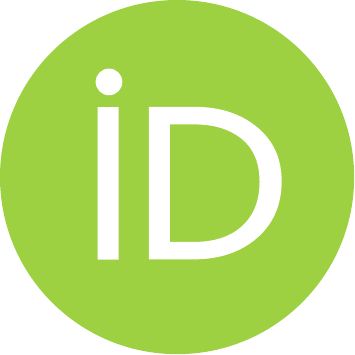}\hspace{1mm}\color{black}Hassan Dashtian}\\
	Computational Media Lab, \\
	School of Journalism and Media,\\
	Moody College of Communication,\\
	The University of Texas at Austin\\
	\texttt{dashtian@utexas.edu} \\
	\And
	\href{https://orcid.org/0000-0001-9734-1124}{\includegraphics[scale=0.06]{orcid.pdf}\hspace{1mm}\color{black}Dhiraj Murthy} \thanks{Corresponding author. E-mail address: \color{blue}Dhiraj.Murthy@austin.utexas.edu \color{black}(Dhiraj Murthy)}  \\
	Computational Media Lab, \\
	School of Journalism and Media,\\
	Moody College of Communication,\\
	The University of Texas at Austin\\
	\texttt{Dhiraj.Murthy@austin.utexas.edu} \\
}

\renewcommand{\shorttitle}{CML-COVID}

\hypersetup{
pdftitle={A template for the arxiv style},
pdfsubject={q-bio.NC, q-bio.QM},
pdfauthor={David S.~Hippocampus, Elias D.~Striatum},
pdfkeywords={First keyword, Second keyword, More},
}

\begin{document}
\maketitle

\begin{abstract}
	As a platform, Twitter has been a significant public space for discussion related to the COVID-19 pandemic. Public social media platforms such as Twitter represent important sites of engagement regarding the pandemic and these data can be used by research teams for social, health, and other research. Understanding public opinion about COVID-19 and how information diffuses in social media is important for governments and research institutions. Twitter is a ubiquitous public platform and, as such, has tremendous utility for understanding public perceptions, behavior, and attitudes related to COVID-19. In this research, we present CML-COVID, a COVID-19 Twitter data set of 19,298,967 million tweets from 5,977,653 unique individuals and summarize some of the attributes of these data. These tweets  were collected between March 2020 and July 2020 using the query terms ‘coronavirus’, ‘covid’ and ‘mask’ related to COVID-19. We use topic modeling,  sentiment analysis, and descriptive statistics to describe the tweets related to COVID-19 we collected  and the geographical location of tweets, where available. We  provide information on how to access our tweet dataset (archived using twarc) at \href{https://doi.org/10.18738/T8/W1CHVU}{\color{blue}https://doi.org/10.18738/T8/W1CHVU}. 
\end{abstract}


\section{Introduction}
COVID-19, an unparalleled global health emergency, led to an exceptional social response on  social media platforms, which includes posts related to social, political and economic life. High volumes of COVID-19-related misinformation  are also present on online social networks such as Twitter [1]. As 68\% of Americans report they use social media to access information and news [2,3], it is critical that understandings of attitudes, perceptions, and responses to COVID-19 are studied using social media data. Furthermore, one third of people report that Twitter is the most important source of scientific information and news [3]. Twitter, on the other hand, can be a source of misinformation about health issues such as vaccination [4]. While the Ebola outbreak in 2014 [5] and the spread of Zika in 2016 [6] highlight the importance of studying pandemics in the content of social networks [7,8], there is a new urgency in monitoring social media content related to COVID-19 [3]. Specifically, social media data related to the COVID-19 pandemic  can, for example, be used to study: (1) the impact of social networks on health info-/mis-information, (2) how misinformation diffusion and spreading can influence behavior and beliefs and (3) evaluate the effectiveness of COVID-19-related actions and campaigns deployed by agencies and governments at global and local scales [9]. 

In this paper, we explore the frequency of tweet activity related to COVID-19 and we make our data and source code publicly available for others to use. We collected tweets real time using the Twitter API from March - July 2020 with the following COVID-19-related query terms (‘coronavirus’, ‘covid’ and ‘mask’). Here, we describe our data collection methods, present basic statistics  of the dataset, and provide information about how to obtain and use the data. We collected over 19,298,967 million tweets from March – June 2020.

\section{Methods}

\subsection{Data Collection}
Our curated data set, CML-COVID, includes 19,298,967 million tweets from 5,977,653 unique individuals from March – June 2020. An average user  in our dataset posted 3 tweets on average. All data were collected from Twitter through Netlytic \footnote{\href{https://netlytic.org/}{\color{blue}https://netlytic.org/}} [11], which queried the Twitter REST API. The dataset is roughly 15 GB of raw data. To comply with Twitter’s Terms \& Conditions (T\&C), we have not publicly released the full text/API-derived information from the collected tweets. Rather, our released data set includes a list of the tweet IDs that others can use to retrieve the full tweet objects directly from Twitter using their own API calls. There are a variety of tools to accomplish this task such as Hydrator\footnote{\href{https://github.com/DocNow/hydrator}{\color{blue}https://github.com/DocNow/hydrator}}. Twitter also provides documentation in their Developer site\footnote{\href{ https://developer.twitter.com/en/docs/twitter-api/v1/tweets/post-and-engage/api-reference/get-statuses-lookup}{\color{blue}https://developer.twitter.com/en/docs/twitter-api/v1/tweets/post-and-engage/api-reference/get-statuses-lookup}} on how to hydrate 100 tweets per API request. 
\subsection{Preprocessing}
First, we pre-processed each raw tweet by concatenating and converting csv files into Python DataFrames and lists to optimize our subsequent data processing. The pre-processing task includes removing characters such as “$\setminus, \slash, \ast$ and etc.” and filtering out stop words (including most rare and most frequent words), and performing text tokenization. This step is essential to next steps which includes topic modeling and sentiment analysis. For topic modeling we applied an unsupervised topic clustering technique called Latent Dirichlet Allocation (LDA). We used TextBlob\footnote{\href{https://textblob.readthedocs.io/en/dev}{\color{blue}https://textblob.readthedocs.io/en/dev}} to perform sentiment analysis. We found extraneous terms (e.g., ‘amp’, ‘dan’, and ‘na’) in our derived topic models. Therefore, we re-ran LDA and removed these terms to present clearer topic modeling results (see table 3).  

\subsection{Data Summary}
A preliminary analysis of the data shows that English is the dominant language in the tweets we collected (65.4\%). One reason for this is that the keywords that we used for querying the Twitter API were all English-language; however other languages are also notably present. For example, 12.2\% of the tweets are Spanish-language. Table 1 summarizes the top 10 languages, the frequency of associated tweets, and the percentage of each language in our  dataset. 63 different languages were identified among the tweets and 3.4\% of tweets had an undefined language. 

\begin{table}[h!]
\centering
\caption{Top ten most popular languages, the number of associated tweets and their percentage.}
\begin{tabular}{|l|l|l|l|}
\hline
ISO Language Code & Language   & Number of Tweets & Percentage \\ \hline
en                & English    & 12488955         & 65.4\%     \\ \hline
es                & Spanish    & 2333241          & 12.2\%     \\ \hline
pt                & Portuguese & 728483           & 3.8\%      \\ \hline
und               & undefined  & 651141           & 3.4\%      \\ \hline
fr                & French     & 536100           & 2.8\%      \\ \hline
in                & Indonesian & 483566           & 2.5\%      \\ \hline
ja                & Japanese   & 419953           & 2.2\%      \\ \hline
it                & Italian    & 262602           & 1.4\%      \\ \hline
tl                & Tagalog    & 183694           & 1.0\%      \\ \hline
hi                & Hindi      & 155204           & 0.8\%      \\ \hline
\end{tabular}
\end{table}
Also, we summarize the top 10 locations (by country and city) of users in Table 2. We retrieved the location of users based on what is reported in their profiles. Therefore, a user’s location is derived from text in their profiles and not GPS coordinates. Locations such as “USA” and “United States” are considered the same and merged as a singular location (i.e., ‘United States’). 

\begin{table}[h!]
\centering
\caption{Top ten locations of tweets based on the user profiles.}
\begin{tabular}{|l|l|}
\hline
Location              & Number of Tweets \\ \hline
‘ ’ (undefined)       & 5483327          \\ \hline
United States         & 330563           \\ \hline
India                 & 121037           \\ \hline
New York, USA         & 85236            \\ \hline
London, England       & 156034           \\ \hline
Washington, D.C., USA & 79412            \\ \hline
Los Angeles, USA      & 79335            \\ \hline
California, USA       & 73098            \\ \hline
México                & 54689            \\ \hline
United Kingdom        & 53773            \\ \hline
\end{tabular}
\end{table}

For each state in the United States with identifiable state-level location, we counted the number of tweets and calculated the frequency of tweets per day. These are illustrated in Figure 1. The United States has the highest frequency of tweets during the period that we collected these data. The number of tweets are low for most regions and countries. As figure 1 illustrates, Canada, Saudi Arabia and India also have a high volume of tweets.  

\begin{center}
\begin{figure*}[htb]
\centerline{\includegraphics[scale=0.38]{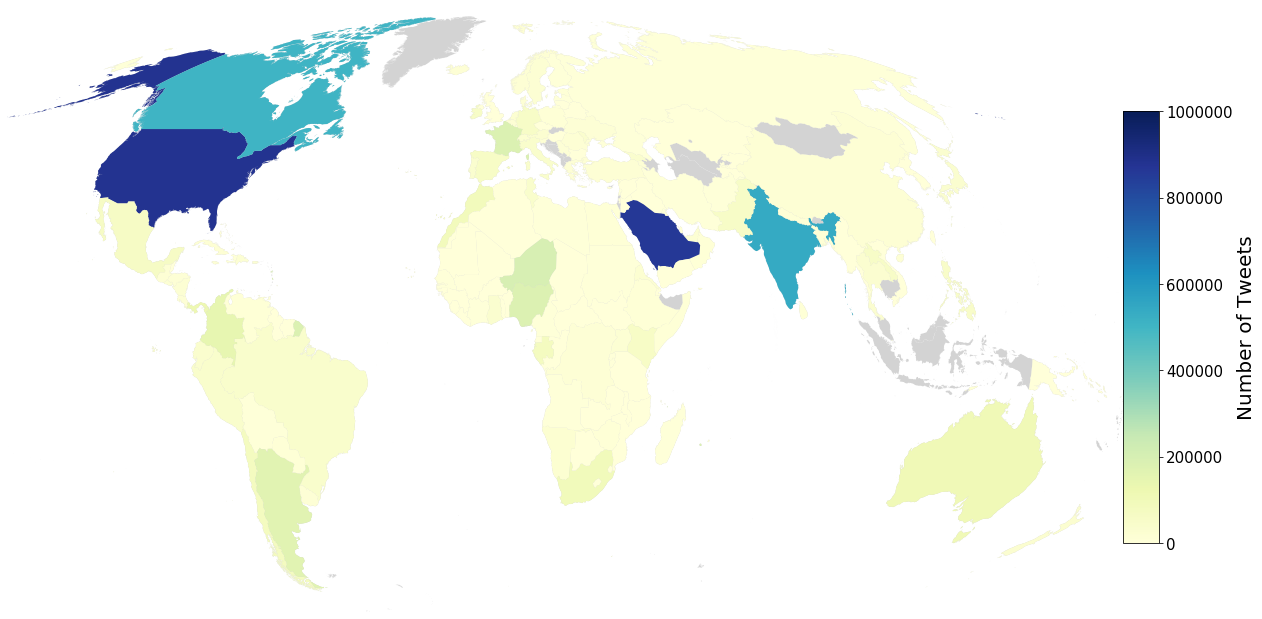}}
\caption{Global distribution, frequency and geographical coverage of the tweets.}
\label{fig:1}
\end{figure*}
\end{center}

Figure 2 depicts the frequency and distribution of tweets in the mainland US. California, New York and Texas have the most number of tweets. The distribution of tweets by state has some similarity to US population distribution.  California (“CA”) appeared in 355,364 of those tweet ID locations,   New York (“NY”) appeared in 295,289 user locations, and  Texas (“TX”)  appeared in 163,920 user locations.  

\begin{center}
\begin{figure*}[htb]
\centerline{\includegraphics[scale=0.38]{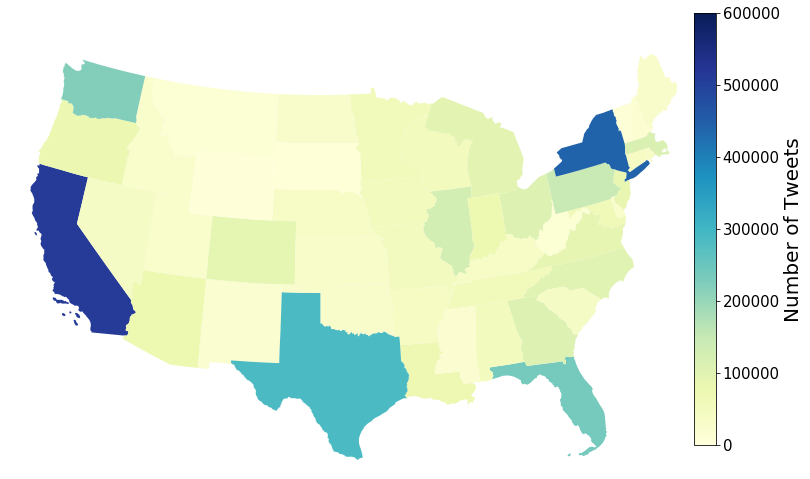}}
\caption{Distribution, frequency and geographical coverage of the tweets in the mainland of US.}
\label{fig:2}
\end{figure*}
\end{center}

We then conducted a frequency analysis by time. We identified the date and time of each tweet and counted the frequencies of tweets for each day as illustrated in Figure 3. Tweet frequency is relatively consistent during our data collection period. We then calculated the sentiment of each tweet. Though sentiment analysis has its limitations with large tweet  corpora, we do believe, like others, that there is some utility in understanding top-level sentiment of these data [10]. To extract information related to sentiment in our collected tweets, we used Textblob to extract the sentiment and scores. We divide tweet sentiment into three main categories -  ‘Negative’, ‘Neutral’ and ‘Positive’. For each day we count the number of tweets with one of these three categories. Figure 4 depicts the time evolution of sentiment by category. As figure 4 indicates, neutral tweets were the most numerous, followed by positive tweets. The gap in frequency between the three sentiment categories is initially, in the first two weeks of April, 2020 reasonably large, but closes after the second week of April, 2020. 

\begin{center}
\begin{figure*}[htb]
\centerline{\includegraphics[scale=0.48]{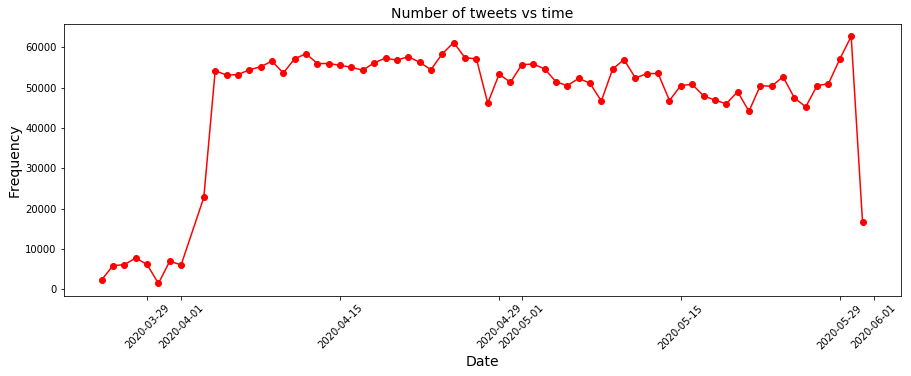}}
\caption{Frequency of tweets related to COVID-19 per day from March to June 2020.}
\label{fig:3}
\end{figure*}
\end{center}

To perform topic modeling, we sampled 20\% of the tweets in our dataset and trained an LDA model that was used to estimate the top most representative words in each topic. Using the trained LDA-based topic model, we obtained 10 topic clusters. Table 3 illustrates the top ten most representative terms associated with each detected ‘topic’ (3 topics are illustrated in table 3; topic 1 is Spanish-language). Since tweets can be in any of 64 different languages, the topics and the top words may contain words and symbols that are from different languages. As we found, cleaning the data based on stopwords in one language is not enough to solve these issues. 
\begin{table}[h!]
\centering
\caption{Examples of three topics, the top ten most representative words and their weights.}
\begin{tabular}{|l|l|l|l|l|l|}
\hline
Topic 0 words & Topic 0 weights & Topic 1 words & Topic 1 weights & Topic 2 words & Topic 2 weights \\ \hline
covid         & 62819.6         & covid         & 82813.9         & covid         & 86130.0         \\ \hline
case          & 13357.3         & coronaviru    & 21099.4         & peopl         & 18270.8         \\ \hline
new           & 10122.1         & \#covid       & 12967.9         & coronavir     & 14549.7         \\ \hline
coronaviru    & 10029.4         & do            & 10428.6         & get           & 13465.0         \\ \hline
test          & 6716.8          & caso          & 8984.3          & like          & 11891.6         \\ \hline
\#covid       & 6676.0          & di            & 8884.3          & death         & 11731.4         \\ \hline
death         & 6579.8          & da            & 8110.9          & go            & 10632.7         \\ \hline
updat         & 4990.1          & si            & 7257.3          & test          & 10174.4         \\ \hline
via           & 4887.8          & \#coronavirus & 6097.0          & one           & 9492.1          \\ \hline
report        & 4828.7          & com           & 5698.3          & us            & 9401.0          \\ \hline
\end{tabular}
\end{table} 
\begin{center}
\begin{figure*}[htb]
\centerline{\includegraphics[scale=0.48]{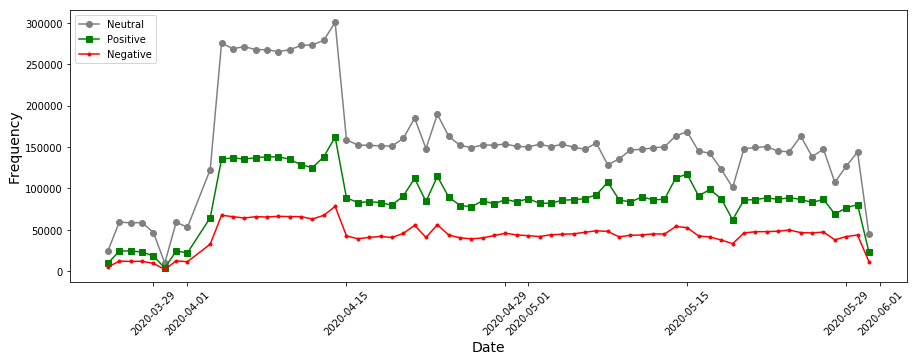}}
\caption{Daily evolution of sentiment of tweets by frequency.}
\label{fig:4}
\end{figure*}
\end{center}

\section{Access}
CML-COVID is a publicly accessible dataset available via the University of Texas Data Repository at the following URL: \href{https://doi.org/10.18738/T8/W1CHVU}{\color{blue}https://doi.org/10.18738/T8/W1CHVU}. It is maintained by the Computational Media Lab (CML; \href{http://www.computationalmedialab.com/}{\color{blue}http://www.computationalmedialab.com/}) in the School of Journalism and Media in the Moody College of Communication at the University of Texas at Austin. The CML-COVID dataset is released in compliance with Twitter’s Terms \& Conditions (T\&C), which prohibit the verbatim release of full tweet text and API-derived data. Rather, we provide a list of tweet IDs that others can directly ‘hydrate’ using calls to the Twitter API. If you use the CML-COVID dataset, please cite this paper or the dataset to acknowledge your use of our results or data. Please contact Hassan Dashtian [{\color{blue}dashtian@utexas.edu}] if you have questions regarding data access and Dhiraj Murthy [{\color{blue}dhiraj.murthy@austin.utexas.edu}] with licensing/usage questions. Source code is also available from the authors and accessible to conduct replication work.

\bibliographystyle{unsrtnat}
\bibliography{references}  
[1] Brennen, J. Scott, et al. "Types, sources, and claims of COVID-19 misinformation." Reuters Institute 7 (2020): 3-1.

[2] Ortiz-Ospina, Esteban. "The rise of social media." Our World in Data 18 (2019).

[3] Singh, Lisa, et al. "A first look at COVID-19 information and misinformation sharing on Twitter." arXiv preprint arXiv:2003.13907 (2020).

[4] Broniatowski, David A., et al. "Weaponized health communication: Twitter bots and Russian trolls amplify the vaccine debate." American journal of public health 108.10 (2018): 1378-1384.

[5] Gomes, Marcelo FC, et al. "Assessing the international spreading risk associated with the 2014 West African Ebola outbreak." PLoS currents 6 (2014).

[6] Petersen, Eskild, et al. "Rapid spread of Zika virus in the Americas-implications for public health preparedness for mass gatherings at the 2016 Brazil Olympic Games." International Journal of Infectious Diseases 44 (2016): 11-15.

[7] Crook, Brittani, et al. "Content analysis of a live CDC Twitter chat during the 2014 Ebola outbreak." Communication Research Reports 33.4 (2016): 349-355.

[8] Fu, King-Wa, et al. "How people react to Zika virus outbreaks on Twitter? A computational content analysis." American journal of infection control 44.12 (2016): 1700-1702.

[9] Cinelli, Matteo, et al. "The covid-19 social media infodemic." arXiv preprint arXiv:2003.05004 (2020).

[10] Kiritchenko, Svetlana, Xiaodan Zhu, and Saif M. Mohammad. "Sentiment analysis of short informal texts." Journal of Artificial Intelligence Research 50 (2014): 723-762.

[11] Gruzd, Anatoliy. "Netlytic: Software for Automated Text and Social Network Analysis." (2016) Available at http://Netlytic.org.






\end{document}